\documentclass[11pt]{article} 
\usepackage{vietnam,epsfig} 
 
\def\mnras{{\em MNRAS}} 
\def\apj{{\em ApJ}} 
\def\apjs{{\em ApJS}} 
\def\aap{{\em A\&A}} 
\def\prd{{\em Phys.~Rev.} D}

\def\be{\begin{equation}} 
\def\ee{\end{equation}} 
\def\bea{\begin{eqnarray}} 
\def\eea{\end{eqnarray}}

\newcommand{\bed}{\begin{displaymath}}
\newcommand{\eed}{\end{displaymath}}
 
\newcommand{\chandra}{\emph{Chandra}}
\newcommand{\xmm}{XMM-\emph{Newton}}
\newcommand{\tspec}{$T_{\rm spec}$}
\newcommand{\tew}{$T_{\rm ew}$}
\newcommand{\tsl}{$T_{\rm sl}$}

\begin{document} 
\vspace*{4cm} 
\title{ SPECTROSCOPIC-LIKE TEMPERATURE OF CLUSTERS OF GALAXIES AND 
 COSMOLOGICAL IMPLICATIONS} 

\author{P.MAZZOTTA$^{1,2}$, E.RASIA$^3$, S.BORGANI$^4$, L.MOSCARDINI$^5$, 
K.DOLAG$^3$, G.TORMEN$^3$} 
 
\address{$^1$Dipartimento di Fisica, Universit\`a Tor Vergata,
via della Ricerca Scientifica 1, I-00133 Roma, Italy\\
$^2$Harvard-Smithsonian Center for Astrophysics, 60 Garden
Street, Cambridge, MA 02138, USA\\
$^3$Dipartimento di Astronomia,
Universit\`a di Padova, vicolo dell'Osservatorio 2, I-35122 Padova,
Italy\\
$^4$Dipartimento di Astronomia dell'Universit\`a di Trieste,
via Tiepolo 11, I-34131 Trieste, Italy\\
$^5$Dipartimento di Astronomia, Universit\`a di Bologna,
via Ranzani 1, I-40127 Bologna, Italy\\
}     
 
\maketitle 

\abstracts{ 
The thermal properties of hydrodynamical simulations of galaxy
clusters are usually compared to observations by relying on the
emission-weighted temperature $T_{\rm ew}$, instead of on the
spectroscopic X-ray temperature $T_{\rm spec}$, which is obtained by
actual observational data. Here we show that, if the intra-cluster
medium is thermally complex, $T_{\rm ew}$ fails at reproducing $T_{\rm
spec}$.  We propose a new formula, the spectroscopic-like temperature,
$T_{\rm sl}$, which approximates $T_{\rm spec}$ better than a few per
cent. By analyzing a set of hydrodynamical simulations of galaxy
clusters, we also find that $T_{\rm sl}$ is lower than $T_{\rm ew}$ by
20--30 per cent.  As a consequence, the normalization of the
$M$--$T_{\rm sl}$ relation from the simulations is larger than the
observed one by about 50 per cent. If masses in simulated clusters are
estimated by following the same assumptions of hydrostatic equilibrium
and $\beta$--model gas density profile, as often done for observed
clusters, then the $M$--$T$ relation decreases by about 40 per cent,
and significantly reduces its scatter. Based on this result, we
conclude that using the observed $M$--$T$ relation to infer the
amplitude of the power spectrum from the X--ray temperature function
could bias low $\sigma_8$ by 10-20 per cent. This may alleviate the
tension between the value of $\sigma_8$ inferred from the cluster
number density and those from cosmic microwave background and large
scale structure. }

\section{Introduction}

Clusters of galaxies represent a powerful tool for investigating
cosmological parameters like the normalization of the primordial power
spectrum, usually expressed in terms of $\sigma_8$ (e.g. Rosati et
al. 2002). The fundamental quantity entering theoretical models is the
cluster mass, $M$, but its high-precision measurement is still precluded by
the presence of different systematic effects.

One way to derive $M$ comes indirectly from the X-ray temperature $T$,
which is related to the cluster mass. In fact simple theoretical
arguments suggest for virialized systems the existence a $M$--$T$
scaling law.  The $M$--$T$ relation is one of the key ingredients in
the recipes to extract cosmological parameters from the cluster
temperature function (XTF) which also requires a proper comparison
between simulated and observed data.  Although simple in principle,
this comparison is not always trivial. Projection effects and/or
instrumental factors introduce systematic biases that complicate the
correct interpretation of the data.  To overcome this difficulty we
built the X-ray MAp Simulator (X-MAS, Gardini et al. 2004), a software
package devoted to simulate X-ray observations of galaxy clusters
obtained from hydro-N-body simulations.  Using X-MAS we showed that,
if the cluster is highly thermally inhomogeneous, then the projected
spectroscopic temperature $T_{\rm spec}$, obtained from the fit of the
observed spectrum, is significantly lower than the emission-weighted
temperature $T_{\rm ew}$, which may be directly inferred from the
simulations by weighting each particle temperature by its emission
measure.  Depending on the thermal complexity of the clusters, this
discrepancy may affect the normalization of the $M$--$T$ relation and,
consequently, the amplitude of the power spectrum $\sigma_8$ derived
from the XTF.

In Mazzotta et al. (2004) we studied in detail the problem of
performing a proper comparison between temperatures obtained from
X-ray observations and simulations.  We found that although in
principle this comparison requires the actual simulation of the
spectral properties of the clusters, when the cluster temperature is
higher than $T> 2-3$~keV, it is possible to define a new
``spectroscopic-like temperature'' function $T_{\rm sl}$, that
approximates $T_{\rm spec}$ to better that few percent.  In this paper
we highlight the main aspects connected with the derivation of \tsl ~
and, thanks to a set of hydrodynamical simulations of galaxy clusters,
we discuss the implications for the $M$--$T$ relation and for the
determination of $\sigma_8$.

\begin{figure}[t]
\begin{center}
\psfig{file=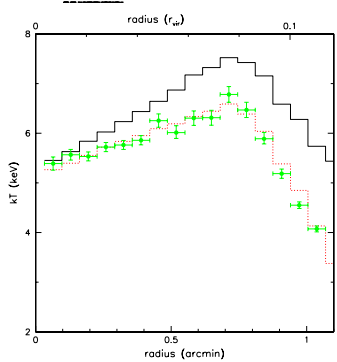,width=5.cm} 
\caption{Temperature profiles of the simulated  galaxy cluster. 
Filled circles indicate the spectroscopic temperature profile from
Gardini et al. (2003); error bars are at $68$ per cent confidence
level for one interesting parameter.  Solid and dotted histogram refer
to the mean emission-weighted and spectroscopic-like temperature
profiles directly derived from the simulation for the same annulus
regions.  }
\label{fig:profiles}
\end{center}
\end{figure}

\section{Spectroscopic-like temperature}

From a purely analytic point of view the spectroscopic temperature
\tspec ~ is not a well defined quantity.  In fact, it results from the
fit of a single-temperature thermal model to a multi-temperature
source spectrum; however, since the former cannot accurately reproduce
the spectral properties of the latter, it follows that \tspec ~ cannot
be unequivocally identified.  Despite of this, as observed spectra are
affected by a number of factors that distort and confuse some of their
properties, in some circumstances multi-temperature thermal source
spectra may appear statistically indistinguishable from
single-temperature ones.  In Mazzotta et al. (2004) we discuss this
aspect by focusing our attention on observations of clusters of galaxies
made using the CCD detectors of \chandra~ and \xmm .  From this study
we find two very important results that we summarize as:
\newline
{\bf i)} given a multi-temperature source spectrum, if the lowest
dominant temperature component has $T_1> 2-3$~keV, then a fit made
with a single-temperature thermal model is statistically acceptable
regardless of the actual spread of the temperature distribution of the
source;
\newline
{\bf ii)} the emission-weighted temperature \tew ~ does not properly
estimate the spectroscopic temperature \tspec ~ as it tends to
overestimate it. The larger the spread in temperature of the thermal
components of the observed spectra, the larger the discrepancy.

To partially solve this problem we propose to use the
spectroscopic-like temperature function \tsl ~instead of \tew .  The
idea behind the analytic derivation of \tsl ~ is quite simple and is
itemized below.

\begin{itemize}
\item Given a multi-temperature source spectrum we are interested in
  identifying the temperature $T_{\rm sl}$ whose spectrum is closest
  to the observed one. If for simplicity we consider a source spectrum
  composed by two thermal models with densities $n_1$, $n_2$ and
  temperatures $T_1$, $T_2$ respectively, requiring matching spectra means that:
\be
\begin{array}{l}
n_{1}^2\zeta(Z,T_1)\frac{1}{\sqrt T_1}\exp(-\frac{E}{kT_1})+
n_{2}^2\zeta(Z,T_2,)\frac{1}{\sqrt T_2}\exp(-\frac{E}{kT_2})\\
\approx A\zeta(Z,T_{sl})\frac{1}{\sqrt T_{sl}}\exp(-\frac{E}{kT_{sl}}),\\
\end{array}
\label{eq:condizion}
\ee 
where $A$ is an arbitrary normalization constant and $\zeta(Z,T)$ is a
parameterization function accounting for the total Gaunt factor and
partly for the line emission. We approximate $\zeta(Z,T,E) \propto
(T/{\rm keV})^{\alpha}$.

\item Supported by the fact that both \chandra\, and \xmm\, are more sensitive
  to the soft region of the X-ray spectrum, we can expand the previous
  equation in Taylor series to the first order of $E/kT$ and obtain:

\be 
T_{\rm sl}= 
\frac{\sum_i n_i^2 T_{i}^\alpha/T_i^{1/2}} 
{\sum_i n_i^2 T_i^\alpha/T_1^{3/2}},
\label{eq:new_form-2} 
\ee  
where we generalized to an arbitrary number of thermal components for
the source spectrum.

\item The power index $\alpha$ needs to be calibrated to the specific
  detector. This can be done by simulating a set of two-temperature
  source spectra and minimizing the value of the mean percentile
  variation of $T_{\rm sl}$ with respect to $T_{\rm spec}$.  In
  Mazzotta et al. (2004) we show that $\alpha=3/4$ represent a good
  choice for each CCD detector of \chandra ~ and \xmm .

\end{itemize}

It is important to say that we tested \tsl ~ against
\tspec , and \tew ~ by  using the hydro-N-body simulation of a thermally
complex (major merger) cluster of galaxies and by applying X-MAS.  We
compared the projected temperature maps and profiles derived by using
different temperature definitions. We find that \tsl ~ approximate
\tspec ~ to better than few percent, while  \tew ~  may
overestimate \tspec ~ by factors as large as 2 (see Mazzotta et
al. 2004 for details).  Just as an example in Fig.~\ref{fig:profiles}
we show the azimuthally averaged temperature profiles of this cluster
as obtained by adopting the three different estimators. It is worth
noticing that while \tsl ~ is consistent within the errors with \tspec
, \tew~ is always higher.
  
\section{Cosmological implications}

In this section we discuss the consequences of the discrepancy between
\tspec ~ and \tew ~ for the $M$--$T$ relation and for the estimate of
$\sigma_8$ from the XTF. To do that we present the analysis of a
sample of hydrodynamical simulations of galaxy clusters for which we
use the temperature function \tsl ~ defined in the previous paragraph.
The results presented here are extracted from Rasia et al. (2005) to
which we refer for more details.

We created the sample of simulated galaxy clusters by merging two
different sets:
\begin{itemize}
\item  Set\#1: 95 temperature-selected clusters (with $T_{\rm ew}>
2$ keV), extracted from the large-scale cosmological hydro-N-body
simulation of a ``concordance'' $\Lambda$CDM model ($\Omega_{0m}=0.3$,
$\Omega_{0\Lambda}=0.7$, $\Omega_{0\rm b}=0.019\,h^{-2}$, $h=0.7$,
$\sigma_8=0.8$), presented in Borgani et al. (2004, B04)

\item  to account for the high temperature ($T>7$~keV) clusters not
  present in Set\#1 because of the limited box size of the previous
  simulation, we added a Set\#2 composed by clusters having $M_{\rm
  vir}>10^{15} h^{-1} M_\odot$ belonging to a different set of
  high-resolution hydro-N-body simulations (Dolag et al. 2005 in
  preparation).
\end{itemize}
Besides gravity and hydrodynamics, both simulations includes the
treatment of radiative cooling, the effect of a uniform
time--dependent UV background, a sub--resolution model for star
formation from a multiphase interstellar medium, as well as galactic
winds powered by SN explosions (Springel \& Hernquist 2003).

\begin{figure}[t]  
\begin{minipage}[b]{0.48\linewidth}
\begin{center}  
\psfig{file=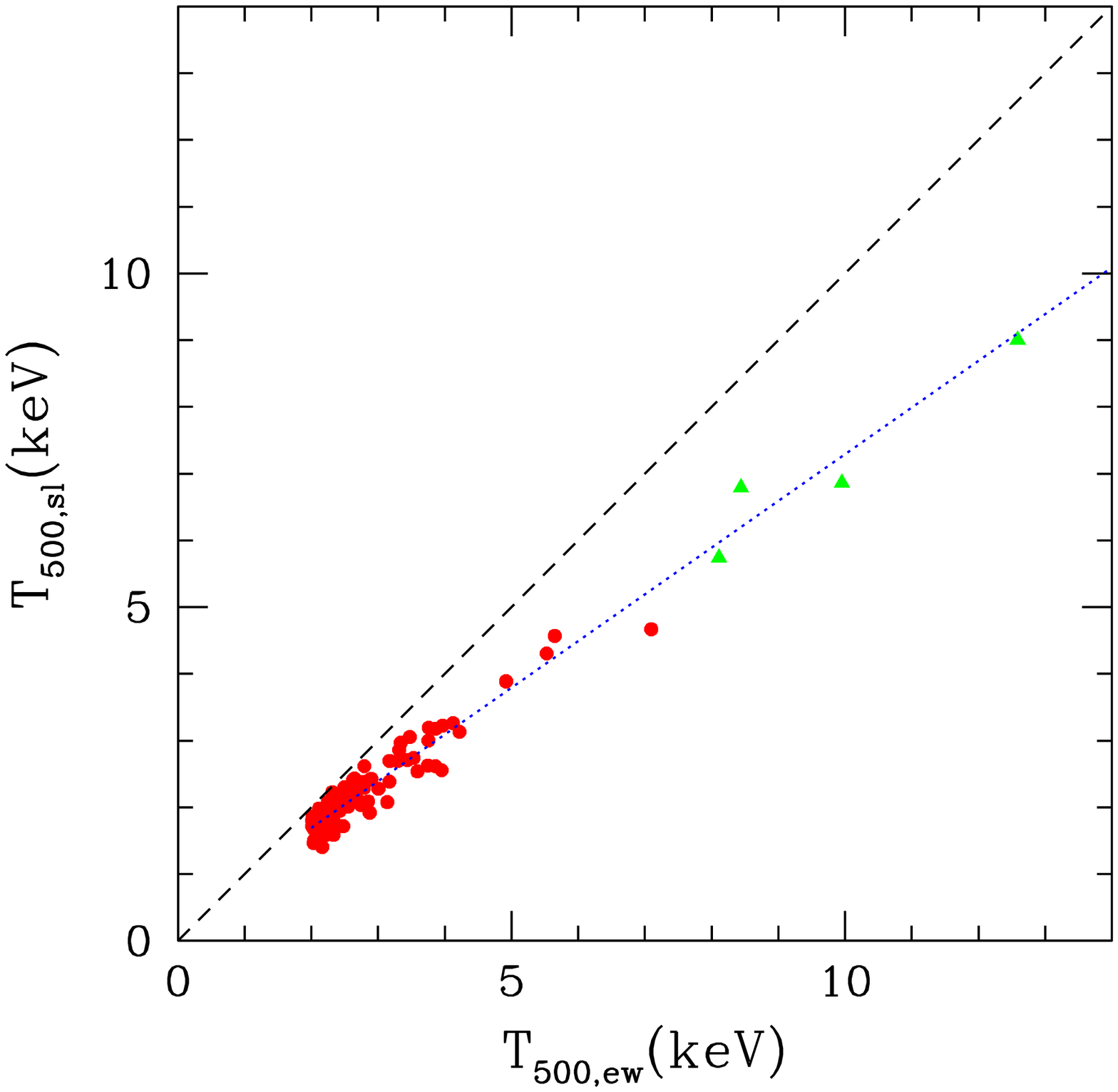,width=5cm}
\caption{The comparison between  $T_{\rm ew}$ and $T_{\rm sl}$. Circles 
and triangles refer to objects from Set\#1 and Set\#2,
respectively. The dotted line is the best fit, while the dashed line
corresponds to $T_{\rm sl}=T_{\rm ew}$.}
\label{fig:TvsT}  
\end{center} 
\end{minipage}
\hspace{0.5cm} 
\begin{minipage}[b]{0.48\linewidth}
\begin{center}  
\psfig{file=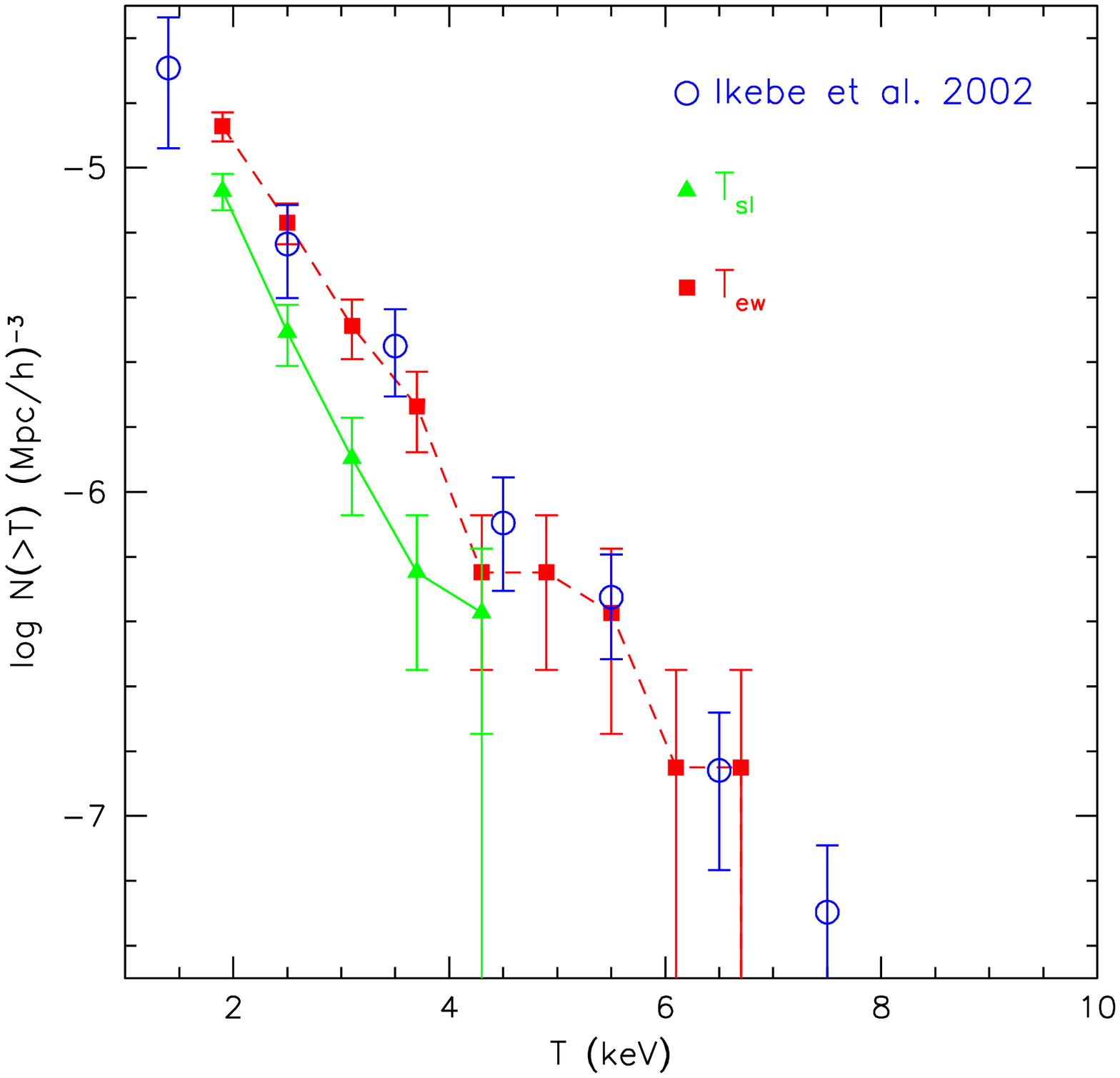,width=5cm}
\caption{The cumulative temperature function  $N(>T)$, 
from Set\#1. Filled squares and dashed line refer to the results for
$T_{\rm ew}$, while filled triangles and solid line represent $T_{\rm
sl}$. Error bars correspond to Poissonian uncertainties. For
reference, open circles show the local temperature function measured
by Ikebe et al. (2002). }
\label{fig:cumul}  
\end{center}
\end{minipage}
\end{figure}

For each cluster in our sample we calculate both $T_{\rm ew}$ and
$T_{\rm sl}$.  In Fig.~\ref{fig:TvsT} we plot the values of $T_{\rm
sl}$ versus $T_{\rm ew}$.  The distribution of the points in
Fig.~\ref{fig:TvsT} suggests that the two estimators can be roughly
related by a linear relation.  By applying a $\chi^2$ fit, we find
$T_{\rm sl}= (0.70\pm 0.01) T_{\rm ew} + (0.29\pm 0.05)$, shown as
dotted line in Fig.~\ref{fig:TvsT}.

The existence of a systematic trend between the two different
estimates of the cluster temperature has interesting consequences on
the $M$--$T$ scaling relation.  Using Set$\#1$, Borgani et al. (2004)
showed that the normalization ($M_0$) of the simulated $M$-$T_{\rm
ew}$ relation is 20\% higher than the observed one.  Since $T_{\rm
ew}$ tends to overestimate $T_{\rm sl}$ by a fair amount, we expect
the actual $M$-$T_{\rm sl}$ relation from simulations to be even more
discrepant with respect to observations. In fact, we find that $M_0$
is higher than the observed one by about 50 per cent.  However, in
order to compare the $M$-$T$ obtained in observations and simulations,
it is also necessary to derive the simulated clusters mass under the
same assumption of the X-ray analysis, {\em i.e.} by applying the
condition of hydrostatic equilibrium to a spherical gas distribution
described by a $\beta$--model (Cavaliere \& Fusco--Femiano 1976), with
the equation of state having a polytropic form.  In this way, the
total self--gravitating mass within the radius $r$ is given by
(Finoguenov et al. 2001; Ettori et al. 2002),
\be 
M(<r)\,=\,1.11\times10^{14} \beta_{\rm fit}\gamma {T(r)\over {\rm
keV}} {r \over h^{-1}Mpc} {x^2\over 1+x^2}\, h^{-1}M_{\odot}.
\label{eq:hyeq} 
\ee   
Here $T(r)$ is the temperature at $r$, $\beta_{\rm fit}$ is the fitted
slope of the gas density profile, $x\equiv r/r_c$ is the radial
coordinate in units of the core radius $r_c$, and $\gamma$ is the
effective polytropic index obtained from temperature and gas profiles.
Fitting the simulation results gives $\log (M_0/h^{-1}M_\odot)= 13.41
\pm 0.13$ when fixing $\alpha=1.5$.
 
As already mentioned, the $M$--$T$ relation is one of the key
ingredients in the recipes to extract cosmological parameters from the
cluster XTF: under the reasonable assumption that the scaling
relations do not significantly change with $\sigma_8$, a larger $M_0$
implies a larger mass for a fixed temperature and, therefore, a higher
normalization of the power spectrum (e.g., Borgani et al. 2001; Seljak
2002; Pierpaoli et al. 2003; Henry 2004).  Huterer \& White (2002)
suggested an approximation for the scaling of $\sigma_8$ with $M_0$,
involving the matter density parameter:
$\Omega_{0m}^{0.6}\sigma_8\propto M_0^{0.53}$.  Based on this relation
and on the fact that a number of analyses of hydrodynamical
simulations (e.g., Muanwong et al. 2002; B04; Rasia et al. 2004; Kay
et al. 2004) have shown that Eq.\ref{eq:hyeq} underestimates the
actual cluster mass by about 20\%, we expect that the value of
$\sigma_8$ is underestimated by a similar amount, {\em i.e.} about
20\%.
 
In Fig.\ref{fig:cumul} we show the effect on the XTF of using $T_{\rm
sl}$, instead of $T_{\rm ew}$. We compare here the cumulative
temperature function $N(>T)$ from the simulation (Set\#1 only) to the
data by Ikebe et al. (2002).  The squared points represent the
emission-weighted temperature; being $T_{\rm ew}$ in good agreement
with the data, one could argue that the $\sigma_8$ used in the
simulation well represents the observed universe. This would imply
$\sigma_8=0.8$.  However, as already shown, the emission-weighted
temperature is typically an underestimate of the observed temperature,
while the spectroscopic-like temperature better takes into account the
observational procedure of temperature measurement.  The comparison
between $T_{\rm sl}$ and data shows that $\sigma_8=0.8$ is excluded at
3$\sigma$ level, while favoring a larger power spectrum normalization,
at least for the ICM physics included in our simulation.  Computing
the mass function by Sheth \& Tormen (1999) for our cosmology at the
mass corresponding to a 2 keV cluster, we find that $\sigma_8$ needs
to be increased to about 0.9 to increase the number density of
clusters above this mass limit by the required 50--60 per cent.

\section{Conclusions}

In this section we simply summarize the main results presented in this
paper.

\begin{enumerate}

\item The projected spectroscopic temperature \tspec ~ of thermally
 complex clusters obtained from X-ray observations is always lower
 than the emission-weighed temperature \tew , which is instead widely
 used in the analysis of numerical simulations.  We show that this
 temperature bias is mainly related to the fact that the
 emission-weighted temperature does not reflect the actual spectral
 properties of the observed source.

\item A proper comparison between simulations and observations needs
 the actual simulations of spectral properties of the simulated
 clusters. Nevertheless, if the cluster temperatures is $>2-3$~keV it
 is possible to define a temperature function, that we call
 spectroscopic-like temperature \tsl , which approximate \tspec ~ to
 better than few per cent.

\item 
By analyzing a sample of hydrodynamical simulations of galaxy
clusters, we find that $T_{\rm sl}$ (and, thus \tspec ) is lower than
$T_{\rm ew}$ by 20--30 per cent.  We obtain a linear fit approximating
the relation between the two different temperature estimators.  As
previous studies made using $T_{\rm ew}$ show that the discrepancy in
the $M$--$T$ relation between simulations and observations is about 20
per cent, it is clear that the use of $T_{\rm sl}$ increases this
discrepancy to $\sim 50$ per cent.  Nevertheless, if we assume
hydrostatic equilibrium for the gas density distribution described by
a $\beta$--model with a polytropic equation of state, we know that
masses are underestimated on average by $\sim 40$ per cent. Although
this goes in the direction of substantially reducing the discrepancy
with observational data, this is not sufficient to cancel it.

\item The bias in the $M$--$T$ relation propagates into a bias in
$\sigma_8$, from the XTF. If such a bias is as large as that found in
our simulations, the values of $\sigma_8$ obtained by combining the
local XTF and the observed $M$--$T$ relation are underestimated by
about 15 per cent.

\item The XTF from the simulation is significantly lower when
using $T_{\rm sl}$ instead of $T_{\rm ew}$. A comparison with the
observed XTF indicates that for the ``concordance'' $\Lambda$CDM model
$\sigma_8$ needs to be increased from $\simeq 0.8$ to $\simeq 0.9$.

\end{enumerate}
 
To conclude, the results of this study go in the direction of
alleviating a possible tension between the power--spectrum
normalization obtained from the number density of galaxy clusters and
that arising from the first--year WMAP CMB anisotropies (e.g. Bennett
et al. 2003) and SDSS galaxy power spectrum (e.g. Tegmark et
al. 2004).

\section*{Acknowledgments} 
This work was partially supported by European contract
MERG-CT-2004-510143 and CXC grants GO3-4163X and GO4-5155X.  We thank
A. Diaferio, G. Murante, S. Ettori, V. Springel, L. Tornatore and
P. Tozzi.
 
 \end{document}